\documentclass[11pt,a4paper]{article}
\usepackage{jinstpub}   
\usepackage[utf8]{inputenc}
\usepackage{booktabs}
\usepackage{array}
\usepackage{tikz}
\usetikzlibrary{arrows.meta,positioning,shapes.geometric}
\newcolumntype{L}[1]{>{\raggedright\arraybackslash}p{#1}}
\newcommand{\nInst}{48}
\newcommand{\nMembers}{256}
\newcommand{\nSeats}{72}
\newcommand{\totalPoints}{288.0}
\newcommand{\instMin}{2}
\newcommand{\instMax}{40}
\newcommand{\nPeriods}{3}
\newcommand{\nSeeds}{10}
\newcommand{\nWeeks}{12}
\newcommand{\nSmall}{23}
\newcommand{\smallSize}{3}
\newcommand{\caseSeed}{2026}
\newcommand{\sizeMedian}{4}
\newcommand{\pDay}{0.8}
\newcommand{\pSwing}{1.0}
\newcommand{\pOwl}{1.2}
\newcommand{\blockShifts}{4}
\newcommand{\vDay}{3.2}
\newcommand{\vSwing}{4.0}
\newcommand{\vOwl}{4.8}
\newcommand{\remoteMult}{0.5}
\newcommand{\shareMin}{2.2}
\newcommand{\fillMinDay}{142}
\newcommand{\fillMinOwl}{213}
\newcommand{\restHours}{12}
\newcommand{\wishMult}{1.5}
\newcommand{\wishMinBlocks}{3}
\newcommand{\hotFrac}{0.2}
\newcommand{\hotWeight}{4}
\newcommand{\shiftLean}{3}
\newcommand{\seatsLow}{70}
\newcommand{\seatsHigh}{72}
\newcommand{\satLow}{84}
\newcommand{\satHigh}{95}
\newcommand{\rOneSeatsNB}{52}
\newcommand{\rOneSeatsTol}{56}
\newcommand{\rTwoSeatsNB}{71}
\newcommand{\rTwoSeatsTol}{72}

\newcommand{\fillMaxNB}{178}
\newcommand{\fillMaxTol}{216}
\newcommand{\overTwoNB}{0}
\newcommand{\overTwoTol}{4}
\newcommand{\satMeanNB}{91}
\newcommand{\satMeanTol}{92}
\newcommand{\satMinNB}{29}
\newcommand{\satMinTol}{42}
\newcommand{\belowFirstNB}{28}
\newcommand{\yearFillMinNB}{71}
\newcommand{\yearFillMaxNB}{130}
\newcommand{\yearFillMinTol}{83}
\newcommand{\yearFillMaxTol}{166}

\newcommand{\nConv}{12}
\newcommand{\shareThree}{3.4}
\newcommand{\boundToQuarter}{5}
\newcommand{\boundToTenth}{11}
\newcommand{\periodsPerYear}{4.3}

\newcommand{\satBase}{94}

\newcommand{\devEndBase}{7}
\newcommand{\toTwentyFiveBase}{4}
\newcommand{\toTenBase}{7}

\newcommand{\nInstMed}{52}
\newcommand{\nMembersMed}{300}

\newcommand{\satMed}{91}

\newcommand{\devEndMed}{8}

\newcommand{\toTenMed}{11}

\newcommand{\nInstLarge}{150}
\newcommand{\nMembersLarge}{1063}
\newcommand{\nSeatsLarge}{216}

\newcommand{\shiftersLarge}{3}

\newcommand{\satLarge}{92}

\newcommand{\devEndLarge}{10}
\newcommand{\toTwentyFiveLarge}{5}

\newcommand{\clsTwoN}{15}
\newcommand{\clsTwoYear}{24}
\newcommand{\clsTwoEnd}{7}
\newcommand{\clsTwoShare}{2.2}
\newcommand{\clsThreeN}{8}
\newcommand{\clsThreeYear}{12}
\newcommand{\clsThreeEnd}{5}
\newcommand{\clsThreeShare}{3.4}
\newcommand{\clsFourFiveN}{13}
\newcommand{\clsFourFiveYear}{11}
\newcommand{\clsFourFiveEnd}{3}
\newcommand{\clsFourFiveShare}{4.5}
\newcommand{\clsSixTenN}{9}
\newcommand{\clsSixTenYear}{8}
\newcommand{\clsSixTenEnd}{5}
\newcommand{\clsSixTenShare}{6.8}
\newcommand{\clsElevenUpN}{3}
\newcommand{\clsElevenUpYear}{13}
\newcommand{\clsElevenUpEnd}{4}
\newcommand{\clsElevenUpShare}{15.8}
\newcommand{\yearPeriods}{4}
\newcommand{\clsBigYear}{13}
\newcommand{\toTenMinFit}{7}
\newcommand{\toTenMaxFit}{10}
\newcommand{\toTenMeanFit}{8.4}

\newcommand{\toTenMinNoFit}{7}
\newcommand{\toTenMaxNoFit}{11}
\newcommand{\toTenMeanNoFit}{10.3}

\newcommand{\compSpreadOn}{83}
\newcommand{\compZeroOn}{13.3}
\newcommand{\compSatOn}{90}
\newcommand{\compSpreadOff}{82}
\newcommand{\compZeroOff}{13.4}
\newcommand{\compSatOff}{91}
\newcommand{\compBand}{10}
\newcommand{\compKinds}{owl, weekend, holiday}
\newcommand{\clsPooledN}{7}
\newcommand{\clsPooledShare}{4.5}
\newcommand{\clsPooledYear}{11}
\newcommand{\clsPooledEnd}{4}
\newcommand{\clsPooledLeft}{1}

\title{Proposing, never assigning: fair allocation of shifts in
experimental particle physics collaborations from ranked wishes under
institutional quotas}
\author{Kamal Benslama}
\affiliation{Department of Physics, Drew University,\\ Madison, NJ 07940, U.S.A.}
\emailAdd{kbenslama@drew.edu}

\abstract{Experimental particle physics collaborations staff the operation of
their detectors with shifts, taken in the control room or remotely by
members drawn from dozens of institutions. The established practice for
filling them is a credit system with self sign-up: institutions owe a quota of
credit, and people take the seats they want when booking opens. This paper
describes a method in which seats are instead allocated from ranked wishes
under institutional quotas, the result is issued as offers that each person
accepts or declines, and an independent checker recomputes every rule
before anything is published: the platform proposes, never assigns, and
proves that it kept the rules. The credit and quota model is inherited
from current practice.

Around that core the paper adds a quota rule that keeps indivisible blocks
fair to institutions whose share of a period is smaller than one block,
and four rules that make fairness visible beyond the count: shares of owl,
weekend and holiday blocks, seats shown in the laboratory's clock and the
institution's own, reachable pools for institutions that cannot travel,
and booking order by balance. On a synthetic collaboration of \nInst{}
institutions and \nMembers{} eligible members, the method allocates
\seatsLow{} to \seatsHigh{} of \nSeats{} seats per period in two rounds,
with a mean rank satisfaction of \satLow{} to \satHigh{}\% and no rule
broken; after a year of running every institution of four members or more
is within \clsBigYear\% of its cumulative share, and the smallest
institutions, which one block overshoots by arithmetic, are brought within
the same range by pooling in pairs. The method is stated completely enough
to be reimplemented; a reference implementation exists and is available to
collaborations under licence. It applies as it stands to any scientific
collaboration that staffs a facility in blocks of shifts under
institutional quotas.}


\arxivnumber{2609.19658}

\begin{document}
\maketitle
\flushbottom

\section{Introduction}
\label{sec:intro}

A particle detector that takes data continuously needs shifts staffed at
every hour, in the control room and, in many experiments, at remote
stations, and a collaboration of several hundred people from several dozen
institutions has to decide who takes them. Shifts are one part of a wider
obligation. The large collaborations share a range of operational work
among their institutions in proportion to their authors: ATLAS counts
shifts together with calibration, data-quality monitoring, maintenance and
software as operation tasks~\cite{atlas_how_works}; CMS tracks operational
contributions as experimental physics responsibilities alongside central
shifts~\cite{cms_join_epr}; ALICE, LHCb and Belle~II define service work of
which shifts are one
kind~\cite{alice_service_work_2026,lhcb_constitution_2026,belle2_bylaws}.
Much of that work is done away from the laboratory, and so are some of the
shifts themselves (Sec.~\ref{sec:practice}). For the shifts, the systems
described publicly work alike: a shift carries credit, the credit counts
toward what an institution owes, and members book the shifts they want when
the period
opens~\cite{atlas_otp_2007,cms_join_epr,belle2_shift_policy_2019}.
The quota settles \emph{how much} each institution owes. Who takes which
seat is settled by first come, first served, that is, by speed.

Speed is a poor allocator. The desirable seats go in minutes, the
undesirable ones are left to whoever feels responsible, and an institution
whose members are asleep when booking opens in the laboratory's time zone
is at a permanent disadvantage. Nothing in the credit system addresses
this, because the credit system was never meant to: it accounts for work
after the fact. The point does not depend on where a seat is or on how much
of a member's obligation shifts represent: it holds for any fixed set of
seats that people sign up for, in a control room or at a remote station,
and it matters in proportion to how contested the seats are.

This paper describes a different second step. Members rank the seats they
would take. The platform then gives seats out, always to the institution
furthest from its quota, its best-ranked wish the rules allow, and lets two
people exchange seats when both prefer the other's. Every allocated seat is
an \emph{offer} with a deadline: the person accepts or declines, and an
unanswered offer returns to the pool. Before any offer is issued, a checker
that shares no code with the allocator recomputes every rule from the input
and the result alone. The principle in one sentence: \emph{the platform
proposes, never assigns, and proves that it kept the rules.}

\paragraph{What is inherited.} The currency is the established one. Shifts
carry points that weigh unpleasant hours more; a weekend block is worth the
same as a weekday block; an institution's quota is its share of the
period's points in proportion to its eligible members; what an institution
falls short of, or exceeds, carries into the next period. This is, in its
essentials, the Belle~II model~\cite{belle2_shift_policy_2019}. ATLAS, CMS
and ALICE count shifts within a wider accounting of operational
work~\cite{atlas_how_works,cms_join_epr,alice_service_work_2026}; the
method needs only what those systems share with it, that a shift seat
carries credit toward what an institution owes.
The model is restated in Sec.~\ref{sec:model} only so that the allocation
can be defined on it.

\paragraph{What is new.} Assigning shifts from stated preferences is
established practice in other fields (Sec.~\ref{sec:practice}), and no
ingredient below is a new mechanism. What this paper contributes is their
combination for the setting of a scientific collaboration, where seats are
owed by institutions under quotas and nobody can be ordered to sit one. It
is the second step, in Secs.~\ref{sec:allocation} and~\ref{sec:blocks}:
\begin{enumerate}
  \item allocation from ranked wishes under institutional quotas, with a
    priority rule chosen for explainability rather than optimality, an
    exchange pass, and a second round for institutions still short;
  \item results as offers with deadlines, so that the platform never books
    anyone into anything;
  \item a quota rule for indivisible blocks: an institution receives a block
    only when the block brings it closer to its quota, and the difference
    carries forward, so that an institution whose share of a period is
    smaller than one block is served every few periods and never ends more
    than half a block above what it is owed;
  \item four rules for fairness beyond the count (Sec.~\ref{sec:fairness}):
    composition shares of owl, weekend and holiday blocks; every seat shown
    in the laboratory's clock and the institution's own; reachable pools for
    institutions that cannot travel; and booking order by balance for what
    the allocation leaves free.
\end{enumerate}
A survey of public descriptions of shift systems in large collaborations
(Sec.~\ref{sec:practice}) found none of these in use for shifts; the closest published
work is a preference-driven scheduler for a clinical physics
division~\cite{rosen_2026}, which optimises preferences but has no
institutional quota, no carry-over and no offer step.

\paragraph{Scope.} The method is written for experimental particle physics,
whose shift model it takes as given: blocks of consecutive shifts, credit
by shift type, quotas by institution, a running balance. It allocates shift
seats and nothing else. A seat may be in the control room or remote; the
method does not distinguish them, and a remote seat differs only in the
credit it carries (Sec.~\ref{sec:model}). The other work a collaboration
credits, and how it weighs that work against shifts, are outside the
method, which needs only that seats carry credit toward a quota. It is described
without reference to any particular experiment; the case of
Sec.~\ref{sec:results} is a synthetic collaboration of typical shape. Nothing in
it is specific to particle physics beyond that shift model, so any
scientific collaboration that staffs a facility the same way, in
astroparticle or nuclear physics, at a telescope or a fusion device, can
use it unchanged. A reference implementation exists and has run the
workflow end to end on the synthetic case (Sec.~\ref{sec:implementation});
the paper does not depend on it.

\section{Definitions}
\label{sec:definitions}

Table~\ref{tab:defs} fixes the words used throughout. They are the words
shifters use, restated so that a reader from outside experimental
physics can follow.

\begin{table}[ht]
\centering\small
\begin{tabular}{L{28mm}L{112mm}}
\toprule
Term & Meaning \\
\midrule
Shift & One staffed span of hours, in the control room or at a remote
  station, of a given type (here day, swing, owl, following the
  laboratory's clock). \\
Block & The unit that is booked and credited: the same shift type on
  consecutive days, either Monday to Thursday (four shifts) or Friday to
  Sunday (three shifts). A person books a block, never a single shift. \\
Seat & One position on one block, in person or remote. A block with one
  shifter per shift has one seat; the seats of a period are the things
  allocated. \\
Points & The credit a seat carries, fixed by its shift type (Sec.~\ref{sec:model}). \\
Period & The span over which quotas are set and allocation runs, here \nWeeks{} weeks. \\
Eligible member & A member who counts in the quota denominator, by a rule the
  collaboration adopts (typically doctoral-level scientists and graduate
  students). \\
Share & An institution's fraction of a period's seat points, in proportion
  to its eligible members. \\
Quota & What an institution owes in a period: its share, plus what it was
  short of (or minus what it exceeded) in earlier periods. \\
Wish & One member's request for one seat, with a rank among that member's
  requests. \\
Offer & An allocated seat proposed to a member, standing until a deadline;
  accepted, it becomes a booking. \\
Round & One wish window followed by one allocation. A period has up to two. \\
Unpleasant block & A block of the kinds a collaboration names as such; here
  an owl block, a Friday-to-Sunday block, or a block on a holiday. \\
\bottomrule
\end{tabular}
\caption{Terms used in this paper.}
\label{tab:defs}
\end{table}

\section{The credit and quota model}
\label{sec:model}

The model is stated here in the form the allocation needs. Every quantity
in it is a setting the collaboration adopts; the values quoted are those of
the synthetic case.

\paragraph{Points.} A weekday shift of type $t$ is worth $p_t$ points, with
the evening shift defining the unit: $p_{\mathrm{day}} = \pDay$,
$p_{\mathrm{swing}} = \pSwing$, $p_{\mathrm{owl}} = \pOwl$. A block is worth
$n \cdot p_t$ where $n$ is the number of shifts in the Monday-to-Thursday
block ($n = \blockShifts$), and the Friday-to-Sunday block is worth the same
despite having one shift fewer. This \emph{equal block value} is the weekend
premium: the shifter who gives up a weekend earns a weekday's worth of
credit for three shifts instead of four. Block values are therefore
$\vDay$, $\vSwing$ and $\vOwl$ points for day, swing and owl. A remote seat,
where the collaboration allows one, is worth a fraction of the in-person
value (here $\remoteMult$); a training (shadow) block earns nothing. The
same ledger credits the duties that are not shifts: an on-call week for a
detector subsystem and a week of run coordination each carry a fixed
number of points to the person's institution. Those duties are assigned by
their own rotas, not by the allocation of Sec.~\ref{sec:allocation}, which
concerns seats only; the seat share of Eq.~\eqref{eq:share} therefore
counts seat points, while an institution's overall standing counts
everything.

\paragraph{Shares and quotas.} Let $T$ be the total seat points of a period
and $e_i$ the eligible members of institution $i$, with $E = \sum_i e_i$.
The share of institution $i$ is
\begin{equation}
  s_i = T\,\frac{e_i}{E},
  \label{eq:share}
\end{equation}
and its quota in the period is its share corrected by its running balance,
\begin{equation}
  q_i = s_i + b_i, \qquad b_i = \sum_{\text{earlier periods}} \big(s_i - a_i\big),
  \label{eq:quota}
\end{equation}
where $a_i$ is what the institution actually held (booked and not
cancelled) in each earlier period. A balance carries without limit and
without expiry. Shares sum to $T$ by construction; quotas do not, because
balances do not sum to zero unless every earlier period was filled exactly.

\paragraph{Why the balance matters.} Equation~\eqref{eq:quota} is what makes
indivisible blocks fair over time. The smallest institution of the case,
with \instMin{} eligible members among \nMembers{}, has a share of
$\shareMin$ points in a \nSeats-seat period: less than one block, so no
single period can give it what it is owed. Section~\ref{sec:blocks} shows
how the allocation treats it.

\section{Allocation from ranked wishes}
\label{sec:allocation}

\subsection{The problem}
\label{sec:problem}

A period has a set of seats $S$, each seat $k$ with points $v_k$, a start,
and an end. Institutions $I$ have quotas $q_i$ from Eq.~\eqref{eq:quota}.
Members $M$ each belong to one institution, $i(m)$. During the wish window
each member submits a ranked list of seats, giving a wish set
$W \subseteq M \times S$ with a rank $r_{m,k}$ for each $(m,k) \in W$. An
allocation is a partial function $x: S \to M$.

\paragraph{Hard rules.} An allocation is admissible when
\begin{enumerate}
  \item each seat goes to at most one person, and only to a person who wished
    for it: $(x(k), k) \in W$;
  \item no person holds two seats that overlap or leave less than a minimum
    rest between them (here $\restHours$ hours);
  \item no person holds more than a maximum number of blocks, if the
    collaboration sets one;
  \item no person holds a shifter seat in a week in which they carry a
    coordination duty that the collaboration declares incompatible with it;
  \item no institution ends further above its quota than the quota rule of
    Sec.~\ref{sec:blocks} allows.
\end{enumerate}

\paragraph{What is sought.} Among admissible allocations, one that first
brings every institution as close to its quota as its wishes allow, with
the institution furthest from its quota served first, and then places each
person as high in their own ranking as possible. The measure of the second
goal used throughout is the rank satisfaction of a person who made $n$
wishes and received the seat ranked $r$:
\begin{equation}
  u(r, n) = 1 - \frac{r - 1}{n},
  \label{eq:util}
\end{equation}
so a first choice scores 1 and a last choice $1/n$. An institution's
satisfaction is the mean over the seats it received.

This is not posed as a single objective to be optimised. The two goals are
ordered, the first is a fairness requirement between institutions and the
second a preference between people, and the collaboration must be able to
explain to any member why a seat went where it did. The rule below is
chosen so that it can. At the size of the problem, of order a hundred
seats and two hundred people per period, it runs in milliseconds; an
exact solver would gain little, at the price of an answer nobody can
retrace.

\subsection{The rule}
\label{sec:rule}

The allocation repeats one step until nothing more can be placed:
\begin{quote}
Pick the institution with the lowest fill, $f_i = a_i / q_i$, among those
that still have a feasible wish. Give it the best-ranked wish of any of its
members that the hard rules allow. Among wishes of equal rank, prefer the
block whose points best fit what the institution is still owed, then the
member with the fewest points so far; break remaining ties by a published
random seed.
\end{quote}
Serving the lowest fill first is what makes the rule a fairness rule: an
institution far behind is served before one nearly full, whatever the order
in which wishes arrived. The block-fit choice is meant to help indivisible
blocks settle: an institution owed a little over two points takes a
three-point day block rather than a five-point owl when both are equally
wished for, and the balance it carries is smaller; how much that matters
is measured in Sec.~\ref{sec:periods}. Preferring the member
with the fewest points spreads the load inside an institution. The seed
makes the run reproducible: the same wishes and the same seed always give
the same allocation, so any allocation can be rerun and audited.

\paragraph{Exchange pass.} After the greedy pass, two people exchange seats
when both would move to a seat they ranked higher, the two seats carry
equal points, and both people stay within the hard rules. Equal points keep
every institution's fill unchanged, so the exchange improves the second
goal without touching the first. The pass repeats until no exchange helps,
bounded by a set number of passes.

\paragraph{Second round.} Popular seats go early, and an institution's
wishes can all be taken by others before its quota is reached. After the
first allocation, institutions still short of their quota get a second,
shorter wish window, restricted to the seats still free, and the same rule
allocates them with the seats already held counted. Two rounds are enough
in practice (Sec.~\ref{sec:results}); whatever remains free afterwards
opens to ordinary first-come booking.

\subsection{Offers}
\label{sec:offers}

Nothing is booked by the allocation. Each allocated seat becomes an offer to
the person, standing for a set number of days. Accepting creates the
booking under the ordinary rules; declining returns the seat to the pool;
an unanswered offer expires at its deadline and the institution's
representative is told. A seat under an open offer cannot be booked by
anyone else in the meantime. The offer step is the reason members can wish
for more seats than they will take: over-wishing is what gives the
allocation room to work, and the offer is where the person makes the final
commitment.

\begin{figure}[ht]
\centering
\includegraphics[width=0.8\textwidth]{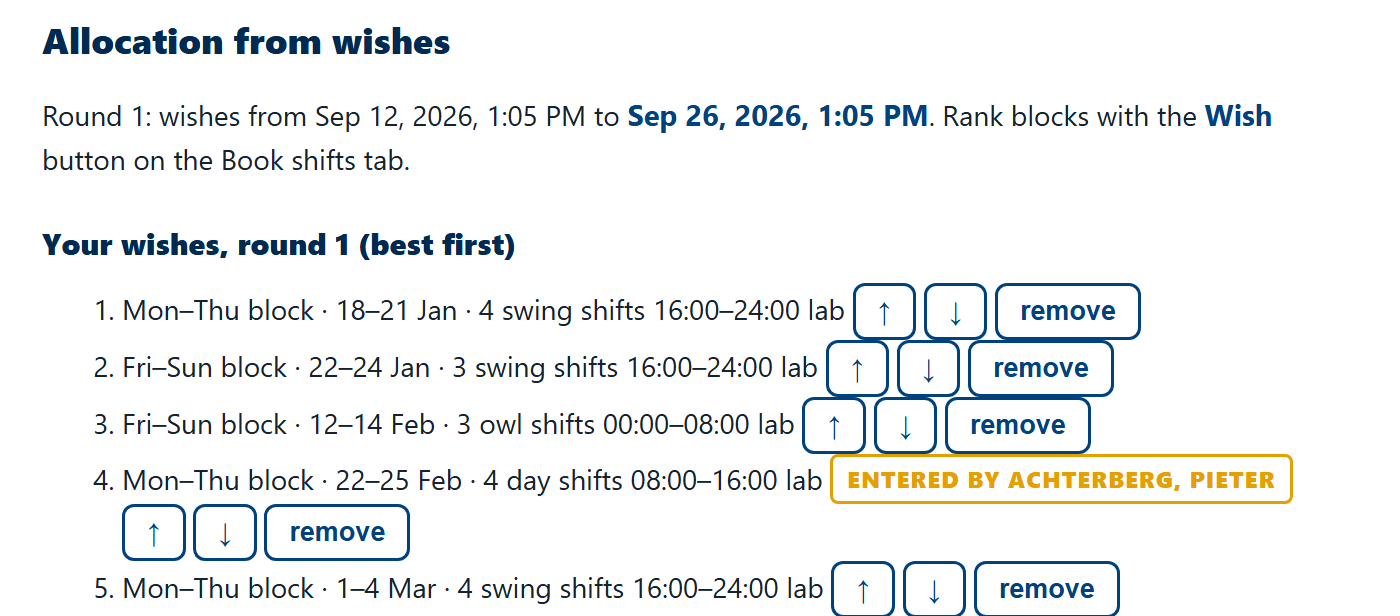}\\[2mm]
\includegraphics[width=0.8\textwidth]{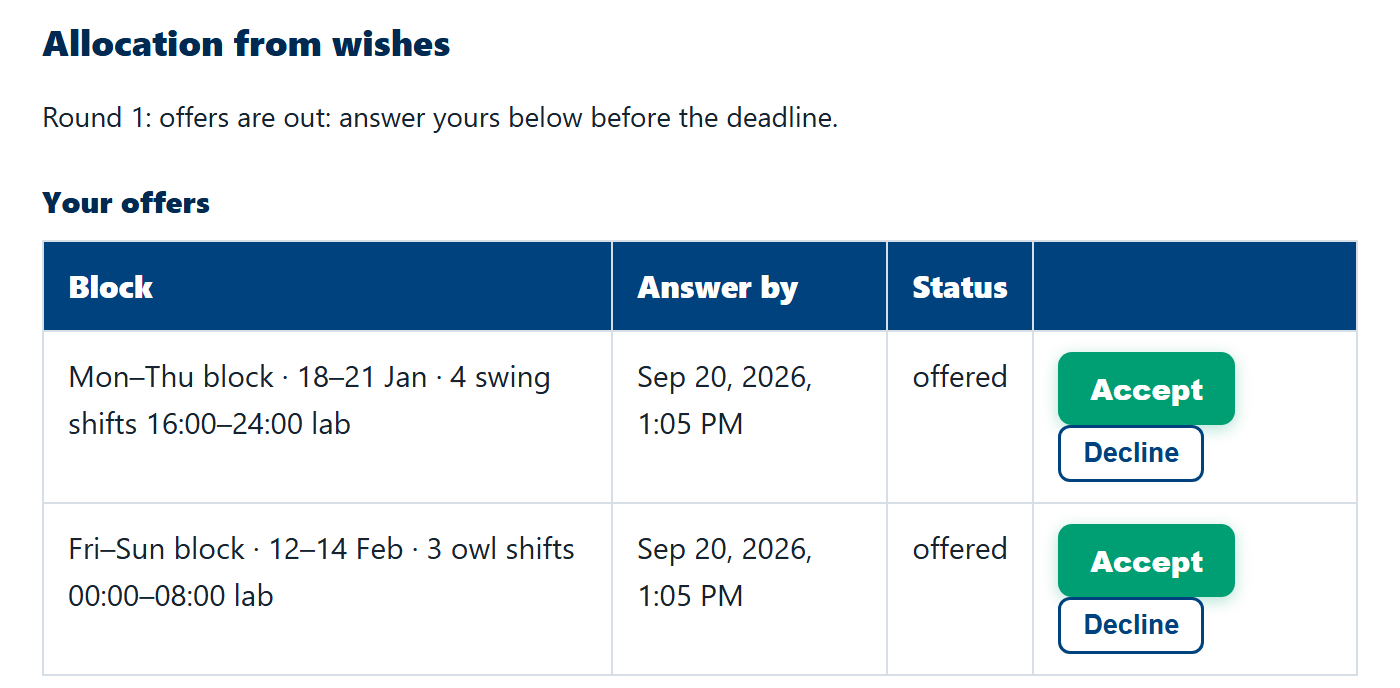}
\caption{A member's view in the reference implementation, on an invented
collaboration. Top: the ranked wishes during the window; the fourth was
entered by the institution's representative and says so. Bottom: the
offers after the allocation, each with its deadline. Every block is named
in full, with its shifts in the laboratory's clock.}
\label{fig:screens}
\end{figure}

Figure~\ref{fig:screens} shows the two moments as a member sees them in
the reference implementation: the ranked list during the window, and the
offers afterwards.

Institutions are asked to wish for at least a multiple of their quota (here
$\wishMult$ times) and for at least a minimum number of blocks (here
$\wishMinBlocks$), whatever their quota. The block minimum matters for
small institutions: a first test with one wish per small institution ended
with those institutions empty-handed whenever someone else took that one
seat.

\subsection{The checker}
\label{sec:checker}

Before offers are issued, a checker recomputes every hard rule from the
input and the proposed allocation alone, with no access to the allocator's
state. It rebuilds each person's holdings seat by seat, verifies the rest
rule and the block limit and the coordination-duty rule, sums each
institution's points against its quota under the quota rule, and confirms
that every allocated seat was wished for. It reports the number of rules
broken, which must be zero, and the per-institution numbers the
collaboration sees: quota, points wished, points allocated, fill,
satisfaction, and share of first choices. In the reference implementation
the same check is run a second time by the database when the result is
published, so that a result that breaks a rule cannot be stored. The
principle is that the allocator may be replaced by any other, including an
exact solver, without changing what is guaranteed.

\section{Indivisible blocks}
\label{sec:blocks}

Blocks cannot be split. An institution whose quota is smaller than a block
therefore cannot land near $100\%$ of it in one period: it receives nothing
($0\%$) or a whole block (for the smallest institution of the case, with a
share of $\shareMin$ points, $\fillMinDay\%$ to $\fillMinOwl\%$). A first
version of the rule allowed any institution to end up to one block above
its quota, which put the smallest institutions well above their share in
every period, with the excess carried as a credit. That is fair over a year
but reads as absurd in a period, and it costs seats for everyone else.

The rule adopted instead is the \emph{nearest-block rule}. An institution
receives a block only when the block brings it closer to its quota, that is
when
\begin{equation}
  a_i + \tfrac{1}{2} v_k \le q_i .
  \label{eq:nearest}
\end{equation}
With Eq.~\eqref{eq:nearest} an institution never ends more than half a
block above its quota. An institution owed less than half the smallest
block in a period receives nothing, is told so, is asked for no wishes, and
carries its share into the next period through Eq.~\eqref{eq:quota}. Two
periods later it is owed about one block and receives one; it then runs
slightly ahead and skips a period; and so on. Over any span its total
never exceeds the sum of its shares by more than half a block. The cost is
a few seats per period that no institution can take under the rule; they
open to first-come booking. The results below quantify both sides.

The rule has no free parameter. The earlier tolerance rule, with its
allowance in points, is kept in the reference implementation as an option
for collaborations that prefer every seat placed by allocation.

\section{Fairness beyond the count}
\label{sec:fairness}

A quota filled to $100\%$ says how much an institution did, not what. Two
institutions at $100\%$ can differ in whether one of them took every
weekend, in whether its members can reach the laboratory at all, and in
whether the night shift at the laboratory is night for them. The
allocation above is fair in the count. The four rules of this section deal
with the composition and the reach, and they are stated here because they
share the ledger of Sec.~\ref{sec:model} and the balance of
Eq.~\eqref{eq:quota}. Each is a setting the collaboration switches on or
off.

\paragraph{Composition shares.} For each institution, its share of all owl
blocks served, of all weekend blocks, and of all holiday blocks, is shown
against its quota share. An institution at $100\%$ of its quota that holds
$5\%$ of the quota and $20\%$ of the weekend blocks is visible as such, and
an institution that holds none of them is equally visible. The shares are
read over the running balance, not per period, so that a group that took
the holidays one period and none the next is not flagged twice. Whether a
share out of line changes anything is the collaboration's decision; the
rule guarantees that it can be seen. The allocation can be asked to act
on it through ties, as an optional addition to the rule of
Sec.~\ref{sec:rule}: among institutions equally far from their quota
(within \compBand\% of fill), the one holding the smallest share of
\compKinds{} blocks is served first, and an institution below its share
prefers such a block among equally ranked wishes. Because only ties are
decided this way, no one is given a seat they ranked lower for it. The
option was built and measured, and the measurement is instructive: on the synthetic case the spread
of unpleasant-block shares across institutions (the standard deviation of
each institution's share of those blocks relative to its quota share) is
\compSpreadOff\% without the priority and \compSpreadOn\% with it, and the
number of served institutions holding none of them \compZeroOff{} against
\compZeroOn{}, at the same satisfaction (\compSatOff\% against
\compSatOn\%). The priority does not move the mix, because the mix is
decided earlier, at the wish stage: an institution whose members never
wish for an owl never receives one, and no tie-break changes that. The
lever that would is a requirement on the wish list, in the same spirit as
the wish multiple, that an institution's wishes include a minimum share
of the unpleasant kinds; or a floor per institution that the checker
verifies. Both are policy. The point weights of Sec.~\ref{sec:model} remain the
only pressure the method applies by itself; the composition shares show
who paid that price, and the remedy, if one is wanted, is the
collaboration's.

\paragraph{Two clocks.} Every seat is defined in the laboratory's clock and
shown, to each viewer, in the clock of the viewer's institution as well.
The institution's time zone is used rather than the person's, so that the
display never reveals where a person is. The rule has a consequence beyond
convenience: an owl shift at a laboratory in the Americas is a working
morning in Europe and a working afternoon in East Asia, so a collaboration
spread over the world can staff its night from other people's day. Where
remote seats exist, showing both clocks turns the least wanted shift at the
laboratory into a normal one for a third of the collaboration, and lets the
allocation's wishes reflect that. The effect is visible in practice: of the
more than 500 CMS online shifts taken in 2025 at a remote operations centre
in the United States, 70\% were the laboratory's night shift, sat in the
local afternoon and evening~\cite{cms_lpc_roc_2026}. No weight is attached to it in this
paper; a collaboration that wishes to price local time can do so as a
setting, but the display alone changes what people wish for.

\paragraph{Reachable pools.} Some institutions have no travel funding and
can serve only from home: remote seats, on-call weeks, coordination duties
that need no presence. Under Eq.~\eqref{eq:share} such an institution owes a
share of seats it can never sit in, ends every period in deficit, and is
marked as behind for something outside its control. The rule adopted is
that a recorded institution is asked only for work it can reach: the period's
points are split into a home pool (everything servable from home), shared
by all institutions in proportion to their members, and an on-site pool,
shared only by the institutions that can travel. The sum is unchanged, so
nothing goes unowed; what a recorded institution cannot take is added to
the others, and each of them sees what was added and why. The price of a
remote seat is the same for everyone; only the cap on how much of a quota
may be earned remotely is lifted for a recorded institution, since the cap
exists to keep the control room staffed, which it cannot do. The status is
recorded in the census, with a reason, and reviewed at every census. None
of the collaborations surveyed in Sec.~\ref{sec:practice} solves this by
asking such institutions for less; the rule here asks them for a different
share of the same total. Credit for service rendered before operations, capped per
institution and in total and added to the period's total so that it
displaces nothing, extends the same ledger and is not described further
here.

\paragraph{Booking order by balance.} Whatever the allocation leaves free
opens to ordinary first-come booking, and so does every period in a
collaboration that does not allocate from wishes. The order in which it
opens is the balance of Eq.~\eqref{eq:quota} at the start of the period:
the institutions that owe the most choose first, in a few tiers a few days
apart, and the first period opens for everyone at once, since there is no
balance yet to sort on. An institution far behind, beyond a set fraction of
its quota, is marked and receives proposals rather than first pick, so
that falling behind is never rewarded with an earlier choice. The ordering
quantity, the tiers, the spacing and the threshold are settings.

Together with the allocation, these are the things a member experiences as
fairness: the count is right, the mix can be seen, the hours are stated
honestly, the demand is one the institution can reach, and what is left
goes to those who are owed the most. None of the four needs a policy vote
beyond the setting that turns it on.

\section{A synthetic case and results}
\label{sec:results}

\subsection{The case}
\label{sec:case}

The case is a synthetic collaboration of the shape that is typical in the
field: \nInst{} institutions and \nMembers{} eligible members, with one
host laboratory of \instMax{} members and a long tail of small groups (the
median institution has \sizeMedian{} eligible members, \nSmall{} have
\smallSize{} or fewer, and none has fewer than \instMin), the sizes drawn
from a log-normal distribution with a fixed seed (\caseSeed). The period is
\nWeeks{} weeks staffed around the clock with one shifter per shift, giving
\nSeats{} seats and \totalPoints{} seat points; shares follow
Eq.~\eqref{eq:share}. Every seat is an in-person seat. A remote seat is a
seat like any other to the allocation, with a lower value and subject to
the cap of Sec.~\ref{sec:model}, so a mixed period changes the numbers but
not the method; it is not simulated here because how many remote seats a
collaboration opens is a policy decision the method does not make. No real person, institution, period or booking
appears, and no number is taken from any real collaboration.

Wishes are generated by the following rules, which are the parameters a
reader needs to regenerate the case:
\begin{itemize}
  \item each institution wishes for at least $\wishMult$ times its share
    and at least $\wishMinBlocks$ blocks, spread at random over its members;
  \item a fraction $\hotFrac$ of the seats are ``popular'': every member is
    $\hotWeight$ times more likely to pick one of them;
  \item each member leans to one shift type, picking it $\shiftLean$ times
    more often than the others;
  \item in the second round, only institutions below their quota wish
    again, among free seats, for at least $\wishMult$ times their gap.
\end{itemize}
Each run uses a seed; the results below are over \nSeeds{} seeds, and every
number is reproduced by the same seed.

\subsection{One period}
\label{sec:oneperiod}

Table~\ref{tab:oneperiod} gives the outcome of one period under the
nearest-block rule and, for comparison, under the tolerance rule with a
one-block allowance.

\begin{table}[ht]
\centering\small
\begin{tabular}{L{62mm}rr}
\toprule
 & Nearest-block rule & Tolerance rule \\
\midrule
Seats allocated, first round & \rOneSeatsNB & \rOneSeatsTol \\
Seats allocated after the second round & \rTwoSeatsNB & \rTwoSeatsTol \\
Highest fill of any institution & \fillMaxNB\% & \fillMaxTol\% \\
Institutions above 200\% of quota & \overTwoNB & \overTwoTol \\
Mean rank satisfaction & \satMeanNB\% & \satMeanTol\% \\
Lowest institution satisfaction & \satMinNB\% & \satMinTol\% \\
Rules broken (checker) & 0 & 0 \\
\bottomrule
\end{tabular}
\caption{One period of \nSeats{} seats, means over \nSeeds{} seeds. Under the
tolerance rule every seat is placed but the smallest institutions end far
above their share; under the nearest-block rule no institution ends more
than half a block above its quota, and a few seats per period open to
first-come booking instead.}
\label{tab:oneperiod}
\end{table}

The second round matters: in the first round \belowFirstNB{} of the
\nInst{} institutions end below quota because their
wishes were taken by others, and the second round, open only to them and
only on free seats, closes most of the gap. The exchange pass rarely fires on this case, because the greedy
order already gives most people a first or second choice; it is kept for
the cases where ranks conflict.

\subsection{Several periods: the number that matters}
\label{sec:periods}

A fill of $\fillMaxNB\%$ in one period is not a failure of the allocation
and not something a better optimiser could remove. A three-member
institution is owed $\shareThree$ points a period; the smallest block is
$\vDay$ and the largest $\vOwl$; whatever the rule, it receives zero, one
or two blocks, and none of those is $100\%$. The number to judge by is the
cumulative fill, the sum of what an institution held over the sum of its
shares, because that is what the balance of Eq.~\eqref{eq:quota} drives to
$100\%$. After \nPeriods{} consecutive periods, under the nearest-block
rule, every institution is between
\yearFillMinNB\% and \yearFillMaxNB\% of the sum of its shares, none more
than half a block above it; under the tolerance rule the spread is
\yearFillMinTol\% to \yearFillMaxTol\%.

\begin{figure}[ht]
\centering
\includegraphics[width=0.62\textwidth]{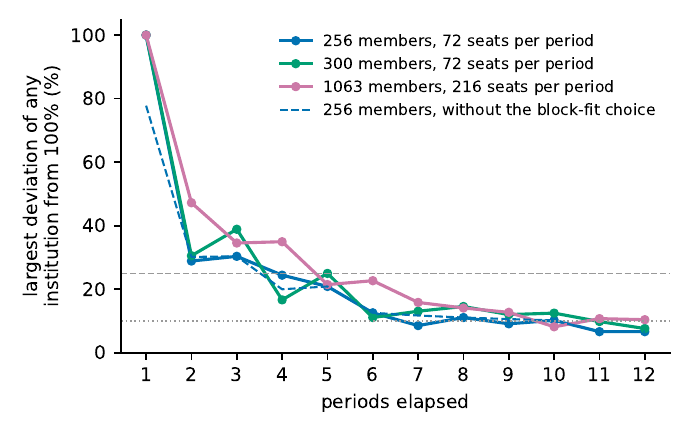}
\caption{The largest deviation of any institution's cumulative fill from
$100\%$, period by period, for three collaboration sizes under the
nearest-block rule, and the first case without the block-fit choice
(dashed blue). Horizontal lines at $25\%$ (dashed) and $10\%$ (dotted).
Synthetic cases, one seed each.}
\label{fig:convergence}
\end{figure}

\paragraph{How fast it converges.} Under the nearest-block rule no
institution is ever more than half a block above its cumulative share, so
its relative overshoot after $k$ periods is bounded by half a block over
$k$ times its share per period, and falls as $1/k$. Below its share an
institution can be further off, for a different reason: its wishes ran
out or clashed with the rules, a shortfall that the second round,
first-come booking and the next period's balance close, and that the
measurements below include. For the smallest
institution of the case, with a share of $\shareMin$ points and a largest
block of $\vOwl$, that bound is $\vOwl / (2 \cdot \shareMin\, k)$: within
$25\%$ after \boundToQuarter{} periods and within $10\%$ after
\boundToTenth{}. Larger institutions converge faster in proportion to their
size, and the measured values below are better than the bound, since a
block rarely overshoots by the full half. With \nWeeks-week periods there
are \periodsPerYear{} periods in a year, so ``within $25\%$ after
\toTwentyFiveBase{} to \toTwentyFiveLarge{} periods'' means within a year of
running, and ``within $10\%$ after \toTenBase{} to \toTenMed{}'' means
between one and a half and two and a half years. For the smallest
institutions the percentage overstates the matter: within $10\%$ of a
$\shareMin$-point share is within a fraction of a point, and the guarantee
that means something to them is the one in points, never more than half a
block from what they are owed.
Figure~\ref{fig:convergence} shows the measured value, the largest deviation
of any institution from $100\%$ of its cumulative share, period by period
over \nConv{} periods, for three collaborations: the case above, a medium
one (\nInstMed{} institutions, \nMembersMed{} eligible members) and a large
one (\nInstLarge{} institutions, \nMembersLarge{} members, \shiftersLarge{}
shifters per shift and \nSeatsLarge{} seats per period). Rank satisfaction
is \satBase\%, \satMed\% and \satLarge\% in the three cases, and the
largest deviation after \nConv{} periods is \devEndBase\%, \devEndMed\%
and \devEndLarge\%. In every case the largest
deviation falls below $25\%$ within \toTwentyFiveBase{} to
\toTwentyFiveLarge{} periods and keeps falling. The first period, taken
alone, is the worst any institution will see; a collaboration that judges
fairness by the per-period fill is judging the wrong number. The dashed
curve shows the case above without the block-fit choice of
Sec.~\ref{sec:rule}. The two curves stay close, which is expected, since
the choice only decides between wishes of equal rank and the half-block
bound is the same. Measured over \nSeeds{} seeds, the
number of periods until every institution is within $10\%$ is
\toTenMeanFit{} on average (\toTenMinFit{} to \toTenMaxFit{}) with the
choice and \toTenMeanNoFit{} (\toTenMinNoFit{} to \toTenMaxNoFit{})
without it: no measurable difference on this case. The choice stays in the rule
because it is the right tie-break and costs nothing, and a case with more
equal-rank ties, for instance institutions whose members wish for the same
blocks, would give it more to decide. On this case the balance settles at
the rate the bound predicts, with or without it.

\paragraph{Which institutions are close after a year, and which are not.}
Table~\ref{tab:classes} answers the question a board will ask: after one
year of running, \yearPeriods{} periods, how far from $100\%$ can an
institution of a given size still be? The answer has a floor and a
shortfall. The half-block bound sets the floor, which shrinks
with the institution's share: after a year it is about $\clsTwoYear\%$ for
a two-member institution and a few percent for a large one. On top of it,
an institution whose wishes ran out or clashed with the rules is short of
seats, and that shortfall does not depend on size; it must be what keeps
the largest class in the table at \clsElevenUpYear\% after a year, since
half a block is only a few percent of its share, and it is closed by the second
round, by first-come booking, and by the balance in the next period. After
a year, then, the two-member institutions are the one class still far from
their share, at \clsTwoYear\%; three-member institutions are at
\clsThreeYear\%; every larger class is within \clsSixTenYear\% to
\clsElevenUpYear\%. After \nConv{} periods every class is within
\clsFourFiveEnd\% to \clsTwoEnd\%. No allocation can move the smallest
institutions faster, because one block is most of their share of a
quarter. A collaboration that finds this unacceptable has two remedies. The
first is to judge the smallest institutions in points rather than in
percent, which is the honest measure for them. The second is
\emph{pooling}: two small institutions agree to be one to the allocation,
with one summed share and one balance, while every seat still credits the
person's own institution. A pool of two two-member institutions has the
share of a four-member one and converges like one. The last row of
Table~\ref{tab:classes} measures it: with the two-member institutions of
the case paired (\clsPooledN{} pools; \clsPooledLeft{} institution without
a partner stays as it is), the pools are within \clsPooledYear\% after a
year, against \clsTwoYear\% unpooled, and \clsPooledEnd\% after \nConv{}
periods. Pooling changes nothing in the
rule; it is a line in the policy saying which institutions may pool and
for how long, and a grouping in the input to the allocation. In the
reference implementation the agreement belongs to the representatives: one
proposes, the others accept, and any of them can end it until the first
wish round of the period opens, after which the pool stands for that
period. The policy line has two values, how small a share must be for an
institution to pool and how many institutions one pool may hold. The round
admits the institutions of a pool together, and the checker verifies the
quota rule for the pool as one unit.

\begin{table}[ht]
\centering\small
\setlength{\tabcolsep}{4pt}
\begin{tabular}{L{34mm}rrrr}
\toprule
Eligible members & Institutions & Share per period (pts) & After \yearPeriods{} periods & After \nConv{} periods \\
\midrule
2 & \clsTwoN & \clsTwoShare & \clsTwoYear\% & \clsTwoEnd\% \\
3 & \clsThreeN & \clsThreeShare & \clsThreeYear\% & \clsThreeEnd\% \\
4 to 5 & \clsFourFiveN & \clsFourFiveShare & \clsFourFiveYear\% & \clsFourFiveEnd\% \\
6 to 10 & \clsSixTenN & \clsSixTenShare & \clsSixTenYear\% & \clsSixTenEnd\% \\
11 and more & \clsElevenUpN & \clsElevenUpShare & \clsElevenUpYear\% & \clsElevenUpEnd\% \\
\midrule
2, pooled in pairs & \clsPooledN{} pools & \clsPooledShare & \clsPooledYear\% & \clsPooledEnd\% \\
\bottomrule
\end{tabular}
\caption{The paper's case by institution size: the largest deviation of any
institution in the class from $100\%$ of its cumulative share, after one
year and after \nConv{} periods. The share is that of the smallest
institution in the class. Last row: the same case with its two-member
institutions pooled in pairs, each pool treated as one institution by the
allocation.}
\label{tab:classes}
\end{table}

\subsection{Sensitivity}
\label{sec:sensitivity}

\begin{figure}[ht]
\centering
\includegraphics[width=0.85\textwidth]{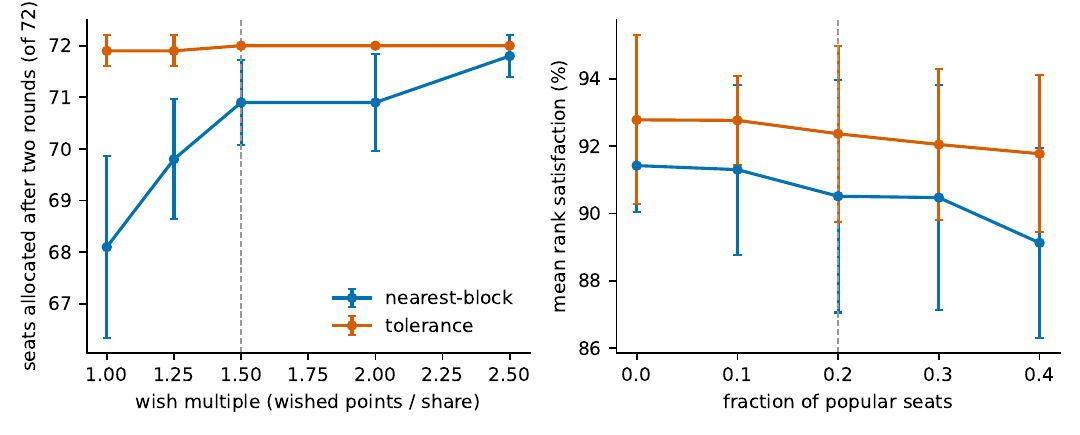}
\caption{Seats allocated after two rounds and mean rank satisfaction, as a
function of the wish multiple (left) and of the fraction of popular seats
(right). Means over \nSeeds{} seeds.}
\label{fig:sensitivity}
\end{figure}

Figure~\ref{fig:sensitivity} sweeps the two parameters a collaboration
would actually set: the wish multiple and the popular-seat fraction. Seats
allocated rise steeply with the wish multiple up to about $1.5$ and flatten
beyond it, which is why that value is proposed. Satisfaction falls slowly
as more seats become popular, which is unavoidable when many people want
the same few seats.

\section{The period in practice}
\label{sec:cycle}

\begin{figure}[ht]
\centering
\resizebox{\textwidth}{!}{%
\begin{tikzpicture}[
  node distance=4mm and 5mm,
  box/.style={draw, rounded corners=2pt, align=center, minimum height=9mm,
              text width=26mm, font=\small, fill=white},
  gate/.style={box, fill=black!6},
  arr/.style={-{Stealth[length=2mm]}, thick},
]
  \node[box] (census) {Census:\\ who counts};
  \node[box, right=of census] (plan) {Run plan:\\ the seats};
  \node[box, right=of plan] (quota) {Quotas:\\ shares + balance};
  \node[box, right=of quota] (wish) {Wish window:\\ ranked lists};
  \node[box, below=of wish] (alloc) {Allocation:\\ lowest fill first};
  \node[gate, left=of alloc] (check) {Checker:\\ 0 rules broken};
  \node[box, left=of check] (offer) {Offers:\\ accept or decline};
  \node[box, left=of offer] (book) {Bookings};
  \draw[arr] (census) -- (plan);
  \draw[arr] (plan) -- (quota);
  \draw[arr] (quota) -- (wish);
  \draw[arr] (wish) -- (alloc);
  \draw[arr] (alloc) -- (check);
  \draw[arr] (check) -- (offer);
  \draw[arr] (offer) -- (book);
  \draw[arr] (offer.south) -- ++(0,-5mm) -| ([xshift=7mm]wish.east)
    node[pos=0.25, below, font=\scriptsize] {round 2: institutions still short wish again}
    -- (wish.east);
  \draw[arr] (book.west) -- ++(-6mm,0) |- (census.west)
    node[pos=0.25, left, font=\scriptsize, align=center] {balance\\carries};
\end{tikzpicture}}
\caption{The period cycle. The allocation proposes; the checker proves the
rules held; people decide by answering offers; the balance carries what
was left into the next period's quotas.}
\label{fig:cycle}
\end{figure}
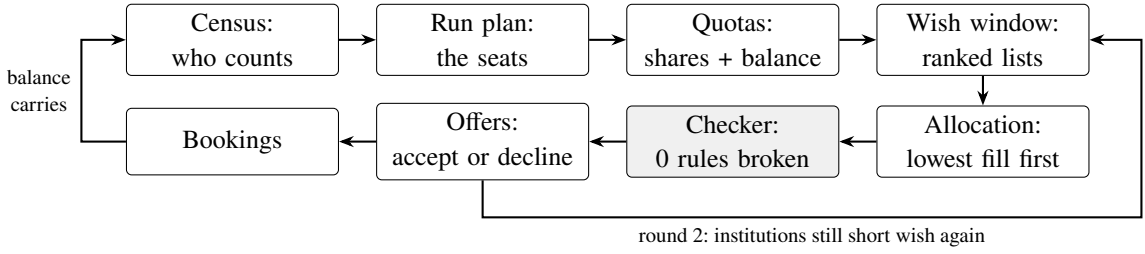

Figure~\ref{fig:cycle} places the allocation in the period cycle. The
census fixes who counts; the run plan fixes the seats; quotas follow from
both by Eqs.~\eqref{eq:share} and~\eqref{eq:quota}; the wish window opens;
the allocation runs and is checked; offers go out and are answered; a
second round serves institutions still short; what remains opens to
ordinary booking; and completed blocks settle the balance that the next
period's quotas carry.

\paragraph{Who does what.} Members rank their own wishes. An institution's
representative sees the institution's list, may add or change entries
(every entry records who made it), and confirms it before the window
closes. The operations coordinators open and close windows, run the
allocation, and publish the offers; they can rerun it with a new seed but
cannot edit its result. An institution at its quota, or waiting under the
nearest-block rule, is left out of a round and told why; nothing is asked
of it. Wishes are visible only to the member, the representative and the
coordinators, since a wish list reveals a person's availability; every
institution's aggregate (points wished, points required, whether
confirmed) is visible to all.

\section{Practice in other collaborations}
\label{sec:practice}

Table~\ref{tab:practice} summarises what public sources describe of the
shift systems of five large collaborations. The survey is limited to what
is published or openly posted; internal tools may do more than their public
descriptions say, and the claim made here is only that no public
description of allocation from ranked wishes with offers was found.

Two things should be read from the table before the comparison. The first
is that shifts are one part of what an institution owes, not the whole.
ATLAS defines operation tasks broadly, as shifts, calibrations,
data-quality monitoring, maintenance and software, and shares them among
its institutions by a fair-share fraction computed from the number of
authors~\cite{atlas_how_works}. CMS institutes owe operational
contributions tracked as experimental physics responsibilities, together
with participation in central shifts, in proportion to their
authors~\cite{cms_join_epr}, and central shifts come in addition to the
general service requirement~\cite{cms_computing_run1_2014}. ALICE
separates shifts in its run control centre and on-call shifts from all
other service work, such as maintenance, calibration, quality control and
data processing, accounts for the two separately, and lets an institution
make up missing service work with more shifts but not the
reverse~\cite{alice_service_work_2026}. The LHCb constitution obliges
members and institutes to take part in service tasks on the detector, its
operation including data-taking shifts, software or
computing~\cite{lhcb_constitution_2026}. Belle~II asks for six months of
service work before authorship and expects institutions to contribute in
proportion to their authors~\cite{belle2_bylaws}. A member of one of these
collaborations can therefore meet much of the obligation without sitting a
control-room shift.

The second is that shifts are no longer only control-room shifts. CMS ran
remote data-quality shifts from centres at Fermilab and DESY during its
cosmic-ray commissioning~\cite{cms_craft_workflows_2010}, built a network
of more than fifty centres used for remote shifts among other
work~\cite{cms_centres_2011}, and in 2025 had more than 500 online shifts
taken at the Fermilab centre by 90 shifters from 34
institutes~\cite{cms_lpc_roc_2026}. ATLAS distributed-computing shifts can
be taken from the shifter's home institute and are credited to that
institute as service~\cite{atlas_adc_shifts_2017}. The Belle~II shift
policy defines remote shifts done from home institutions, with a quota of
their own~\cite{belle2_shift_policy_2019}, and from 2021 its online shift
crew consisted of one shifter on site and two
remote~\cite{belle2_kisti_2023}.

\begin{table}[ht]
\centering\small
\begin{tabular}{L{20mm}L{56mm}L{60mm}}
\toprule
Collaboration & Shifts: how seats are taken and credited & Other credited work; remote shifts \\
\midrule
ATLAS & Self-booking when the window opens; points per task, with night and
  weekend tasks worth about twice a morning; shadow shifts before a first
  shift~\cite{atlas_otp_2007,atlas_otp_classes_mpp_2016}. &
  Operation tasks shared by number of authors: shifts, calibration,
  data-quality monitoring, maintenance,
  software~\cite{atlas_how_works}. Computing shifts from the home
  institute~\cite{atlas_adc_shifts_2017}. \\
CMS & Self-booking of central shifts, shown with the institute's
  expected, pledged and done work~\cite{cms_join_epr,cms_epr_guide_2019}. &
  Experimental physics responsibilities owed by number of
  authors~\cite{cms_join_epr}. Remote shifts from centres at Fermilab, DESY
  and elsewhere~\cite{cms_craft_workflows_2010,cms_centres_2011,cms_lpc_roc_2026}. \\
Belle~II & Self-booking in the shift tool; institution quotas from the
  members database; carry-over across
  years~\cite{belle2_shift_policy_2019}. &
  Six months of service work before authorship~\cite{belle2_bylaws}.
  Separate local and remote shift quotas with no compensation between
  them~\cite{belle2_shift_policy_2019}; remote control-room
  shifters~\cite{belle2_kisti_2023}. \\
ALICE & Rotating blocks of two morning, two afternoon and two night
  shifts; an institute's shift quota follows its number of
  scientists~\cite{alice_sams_2015}. &
  Service work in classes: run-control-centre shifts, on-call shifts, and
  all other service work, accounted
  separately~\cite{alice_service_work_2026}. \\
LHCb & A central crew of shift leader and data manager, with subsystem
  experts on call~\cite{lhcb_run3_2026}. &
  Service tasks on the detector, its operation including data-taking
  shifts, software or computing~\cite{lhcb_constitution_2026}. \\
\bottomrule
\end{tabular}
\caption{Shifts and other credited work in five collaborations, from public
sources. The ATLAS shift entry rests on a CERN note and one institute's
slides; the Belle~II policy document is marked draft; the CMS description
comes from the collaboration's public pages and user guide; the LHCb
entries come from its constitution and a public summary of its last run,
and no public description of how its shift seats are taken was found.}
\label{tab:practice}
\end{table}

Where the public sources say how shift seats are taken, it is by sign-up
with credit accounting, and for the shifts the model of
Sec.~\ref{sec:model} is theirs. None describes an allocation from ranked
wishes with offers. The difference of this paper is in that second step,
and it concerns shift seats only, in the control room or remote. How much
it matters to a collaboration follows from the two observations above.
Where shifts are one of several ways to meet an obligation and there are
many seats of many kinds, the pressure on any one seat is low and the race
at the opening of a booking window is mild. Where shifts are a large part
of what an institution owes, as in a smaller collaboration or an experiment
run from a single site, or where a class of seats is scarce, such as remote
seats at convenient hours, the race is real and the allocation has
something to settle.

Outside physics, preference-integrated scheduling is an
active field: the nurse rostering literature is large and
mature~\cite{burke_nurse_2004,decausmaecker_nurse_2011}, and a recent
clinical physics scheduler asks each physicist to grade every slot and
optimises a penalty over the grades~\cite{rosen_2026}. Those systems assign
people to shifts to cover a demand; they have no institutional quota, no
balance carried between periods, and no offer step, because a hospital
schedule is an instruction, not a proposal. Fairness and acceptance in
collaborative shift scheduling has been studied from the users'
side~\cite{uhde_chi_2020}, with the finding that people accept a schedule
they can see the reasons for, which is the argument for a rule that can be
retraced over one that is optimal.

In the economics of matching, serving institutions by an ordering and
giving each its best remaining choice is a priority
mechanism~\cite{abdulkadiroglu_sonmez_1998,abdulkadiroglu_sonmez_2003},
and the exchange pass is a restricted form of the trading step in such
mechanisms. Multi-unit assignment under quotas, with the analysis of how
far a simple mechanism sits from an efficient one, is treated
in~\cite{budish_2011,budish_cantillon_2012}, and the leximin criterion
(make the worst-off as well off as possible, then the next) that the
lowest-fill-first rule approximates is studied
in~\cite{kurokawa_leximin_2018}. The contribution here is not a new
mechanism in that sense but the combination, in an operating system, of a
quota with carry-over, a priority rule stated so that any member can retrace
it, an independent check, and the offer.

\section{Implementation experience}
\label{sec:implementation}

A reference implementation, \emph{Watchbill}, carries the whole period
cycle of Fig.~\ref{fig:cycle} for a collaboration: the census, the run
plan, quotas, the wish rounds, the allocation and its checker, offers,
booking, swaps, and the rotas for on-call and coordination duties. It has
been exercised end to end, on a case of the same kind and size as that of
Sec.~\ref{sec:case}, by an automated test in which every one of some two
hundred members acts as themselves on their own authenticated session: two
small institutions pool through their representatives before the first
round, the members wish, a representative edits and
confirms, outsiders are refused, the allocation runs, a tampered result is
refused by the second check, offers are accepted and declined, the second
round opens only to institutions still short, and the period is released.
The test ends with every seat holding at most one booking, every booking
before release traceable to an accepted offer, no rest-rule clash, and no
institution or pool above what the quota rule allows; the pool received a
block in a period in which each of its institutions alone would have
waited, and each of its seats is credited to the person's own institution.
A separate test of the
ordinary first-come booking, some two hundred members all attempting to
book at the same instant on a private copy, ended with no seat booked
twice and every loser told at once, the whole rush over in under three
seconds. Two practical lessons came out of this. The allocation itself
takes milliseconds; the time in a period is spent by people, in the wish
window and in answering offers, so the windows have to be set for them,
not for the machine. And the independent check never caught the allocator
out, but during development it caught two mistakes in the logic around it,
which is reason enough to have it.

The implementation is not the subject of this paper and its internals are
not described here. Every value named in this paper is a setting of the
implementation rather than a constant of it, so that a collaboration adopts
the method by choosing values, not by changing code.

The screenshots of Fig.~\ref{fig:screens} are from the reference
implementation running on an invented collaboration; no real name,
institution or booking appears in them.

\section{Limitations}
\label{sec:limitations}

The results are synthetic. The implementation has run the workflow end to
end on the synthetic case and is running as a pilot for one collaboration,
whose policy is proposed and not yet decided; no real period has yet been
allocated from wishes. The first will produce the numbers this paper cannot
yet give, above all how many wishes real members submit and how many offers
are declined. The generating rules of Sec.~\ref{sec:case} are a guess at
that behaviour, chosen to be unkind (a fifth of the seats wanted by
everyone). The rule is greedy and makes no optimality claim; an exact
formulation would be a natural comparison, and the checker is designed so
that any allocator can be substituted. A formal treatment of the rule as a
mechanism, its incentive properties and its distance from an efficient
assignment, belongs to the scheduling and matching literature cited in
Sec.~\ref{sec:practice} and is not attempted here. How much a collaboration
gains from the method depends on how contested its seats are: where shifts
are a small part of what members owe and are easily filled, self sign-up
may serve well enough, and the method earns its place where seats are
scarce, unequal in appeal, or sought from many time zones. Rank satisfaction, Eq.~\eqref{eq:util},
treats ranks as evenly spaced, which they are not for everyone. The
composition priority of Sec.~\ref{sec:rule} acts only through ties, so an
institution that never wishes for an owl never receives one; a hard rule
would be a cap or a floor per person or per institution on unpleasant
blocks, which the checker could verify like the rest rule, and which is
left to a collaboration's policy. Two further extensions are natural and
not built here: a wish for a whole week, granted whole or not at all, for
members who travel far to sit their shifts; and a cap per person on
unpleasant blocks in a year. Finally,
the method allocates seats to institutions' members; it does not decide
what an institution owes when it cannot travel, or how service other than
shifts is credited, which are policy questions the collaboration answers
outside the allocation.

\section{Conclusion}
\label{sec:conclusion}

For shifts, the credit-and-quota model that particle physics collaborations
use settles what each institution owes and leaves who takes which seat to
speed, whether the seat is in the control room or remote. This paper
keeps the model and replaces the race with an allocation from ranked
wishes that serves the institution furthest from its quota first, issues
its result as offers that people accept or decline, and proves by an
independent check that every rule held. A quota rule for indivisible
blocks, with the balance carried between periods, keeps the smallest
institutions fair over a year without giving any of them a multiple of
their share in a period. Four further rules make fairness visible beyond
the count: who took the owls, the weekends and the holidays; what hour a
seat is where the viewer sits; a quota each institution can reach; and a
booking order that serves those owed the most. On a synthetic case of
realistic shape the
method places \seatsLow{} to \seatsHigh{} of \nSeats{} seats per period in
two rounds at a mean rank satisfaction of \satLow{} to \satHigh{}\%, and
never lets an institution run more than half a block above its cumulative
share. The
method is stated fully enough to be reimplemented, and a reference
implementation exists. Any scientific collaboration that staffs a facility
in blocks under institutional quotas can use it as it stands.

\section*{Data and code availability}
This work uses no experimental data. The synthetic collaboration is
specified completely in Sec.~\ref{sec:case}, and the method is stated
completely in Secs.~\ref{sec:model} to~\ref{sec:fairness}, so every result
can be regenerated from the paper alone. The reference implementation is
not publicly archived. It is the author's work, copyright reserved, and is
licensed to scientific collaborations for their operations on request,
with the condition that this paper be cited. Confidential access can be
given to referees.

\section*{Acknowledgements}
The author thanks the colleagues who tested the reference implementation
and whose questions shaped the quota rule for small institutions. No
collaboration's data appear in this paper.

\paragraph{Use of AI-assisted technology.} An AI assistant developed by
Anthropic was used under the author's direction as a drafting and
software-engineering tool: to draft and edit text, and to write and test
code of the reference implementation and of the scripts that generate the
numbers and figures of this paper. The method, the design decisions and
the analysis are the author's, who checked every result and the final text
and takes full responsibility for the content.

\bibliographystyle{JHEP}
\bibliography{references_jinst}   

\providecommand{\href}[2]{#2}\begingroup\raggedright\begin{thebibliography}{10}

\bibitem{atlas_how_works}
{ATLAS Collaboration}, ``How {ATLAS} works.'' ATLAS public web page,
  \url{https://atlas.cern/Discover/Collaboration/How-ATLAS-Works}, Accessed
  2026-11-21.

\bibitem{cms_join_epr}
{CMS Collaboration}, ``How to join {CMS}.'' CMS public web page,
  \url{https://cms.cern/collaboration/how-join-cms}, Accessed 2026-09-15.

\bibitem{alice_service_work_2026}
{ALICE Collaboration}, ``{ALICE} service work, rev. 6.'' ALICE Collaboration
  document,
  \url{https://alice-collaboration.web.cern.ch/sites/default/files/Documents/ALICE_SW_ModusOperandi-Rev6.pdf},
  June, 2026.

\bibitem{lhcb_constitution_2026}
{LHCb Collaboration}, ``The {LHCb} constitution.'' Ratified by the
  Collaboration Board on 24 June 2026,
  \url{https://lhcb.web.cern.ch/lhcb_page/collaboration/organization/constitution/Constitution.pdf},
  2026.

\bibitem{belle2_bylaws}
{Belle II Collaboration}, ``{Belle II} bylaws.'' Adopted 19 June 2017, as
  amended, \url{https://public.belle2.org/downloads/belle2_bylaws.pdf},
  Accessed 2026-11-21.

\bibitem{atlas_otp_2007}
B.~Copy and M.~Tsikanin, \emph{{ATLAS} maintenance and operation management
  system},  Tech. Rep.
  \href{https://cds.cern.ch/record/1055567}{CERN-IT-Note-2007-009}, CERN
  (2007).

\bibitem{belle2_shift_policy_2019}
R.~de~Sangro, ``Belle {II} shift policy document, rev. 2.2.19.'' Belle II
  Collaboration (marked DRAFT),
  \url{https://indico.belle2.org/event/394/contributions/1222/subcontributions/16/attachments/717/1087/Belle2ShiftPolicy2019v2.2.19.pdf},
  May, 2019.

\bibitem{rosen_2026}
B.S.~Rosen, Z.~Zhang, K.M.~Hopfensperger, K.C.~Paradis, C.~Lee, L.~Critchfield
  et~al., \emph{A preference-integrated optimization system for medical physics
  shift scheduling}, \href{https://doi.org/10.1002/acm2.70743}{\emph{Journal of
  Applied Clinical Medical Physics} {\bfseries 27} (2026) }.

\bibitem{cms_lpc_roc_2026}
{Fermilab LHC Physics Center}, ``Shifts at {Fermilab} {LPC} {ROC}.'' Web page,
  updated 7 January 2026, \url{https://lpc.fnal.gov/activities/ROCE/}, Accessed
  2026-11-21.

\bibitem{cms_computing_run1_2014}
J.~Adelman et~al., \emph{{CMS} computing operations during run 1},
  \href{https://doi.org/10.1088/1742-6596/513/3/032040}{\emph{J. Phys.: Conf.
  Ser.} {\bfseries 513} (2014) 032040}.

\bibitem{cms_craft_workflows_2010}
{CMS Collaboration}, \emph{{CMS} data processing workflows during an extended
  cosmic ray run},
  \href{https://doi.org/10.1088/1748-0221/5/03/T03006}{\emph{JINST} {\bfseries
  5} (2010) T03006} [\href{https://arxiv.org/abs/0911.4842}{{\ttfamily
  0911.4842}}].

\bibitem{cms_centres_2011}
L.~Taylor, \emph{{CMS} centres worldwide: a new collaborative infrastructure},
  \href{https://doi.org/10.1088/1742-6596/331/8/082007}{\emph{J. Phys.: Conf.
  Ser.} {\bfseries 331} (2011) 082007}.

\bibitem{atlas_adc_shifts_2017}
C.~Adam, D.~Barberis, S.~Cr\'ep\'e-Renaudin, K.~De, F.~Fassi, A.~Stradling
  et~al., \emph{Computing shifts to monitor {ATLAS} distributed computing
  infrastructure and operations},
  \href{https://doi.org/10.1088/1742-6596/898/9/092004}{\emph{J. Phys.: Conf.
  Ser.} {\bfseries 898} (2017) 092004}.

\bibitem{belle2_kisti_2023}
K.~Kim, K.~Park and K.~Cho, \emph{{KISTI} remote control room for {CDF} and
  {Belle II} collider experiments},
  \href{https://doi.org/10.1007/s40042-023-00763-4}{\emph{J. Korean Phys. Soc.}
  {\bfseries 82} (2023) 638}.

\bibitem{atlas_otp_classes_mpp_2016}
{ATLAS MPP Group}, ``{ATLAS} operation tasks: Responsibilities of the {ATLAS}
  {MPP} group.'' Slides, Max Planck Institute for Physics,
  \url{https://indico.mpp.mpg.de/event/4493/contributions/10569/attachments/8279/9189/OTP_Overview.pdf},
  Oct., 2016.

\bibitem{cms_epr_guide_2019}
{CMS Collaboration, iCMS}, ``{iCMS-EPR} user guide, version 1.1.''
  \url{https://indico.cern.ch/event/896308/contributions/3779326/attachments/2000942/3340046/EPR-UserGuide-v1.1.pdf},
  Nov., 2019.

\bibitem{alice_sams_2015}
H.~Martins~Silva, I.~Abreu Da~Silva, F.~Ronchetti, A.~Telesca, C.~Maidantchik
  and {ALICE Collaboration}, \emph{The {ALICE} glance shift accounting
  management system ({SAMS})},
  \href{https://doi.org/10.1088/1742-6596/664/5/052037}{\emph{J. Phys.: Conf.
  Ser.} {\bfseries 664} (2015) 052037}.

\bibitem{lhcb_run3_2026}
{LHCb Collaboration}, ``Run 3: Sitting on a goldmine.'' LHCb public web page,
  \url{https://lhcb-outreach.web.cern.ch/2026/10/30/end-of-run-3/}, Oct., 2026.

\bibitem{burke_nurse_2004}
E.K.~Burke, P.~De~Causmaecker, G.~Vanden~Berghe and H.~Van~Landeghem, \emph{The
  state of the art of nurse rostering},
  \href{https://doi.org/10.1023/B:JOSH.0000046076.75950.0b}{\emph{Journal of
  Scheduling} {\bfseries 7} (2004) 441}.

\bibitem{decausmaecker_nurse_2011}
P.~De~Causmaecker and G.~Vanden~Berghe, \emph{A categorisation of nurse
  rostering problems},
  \href{https://doi.org/10.1007/s10951-010-0211-z}{\emph{Journal of Scheduling}
  {\bfseries 14} (2011) 3}.

\bibitem{uhde_chi_2020}
A.~Uhde, N.~Schlicker, D.P.~Wallach and M.~Hassenzahl, \emph{Fairness and
  decision-making in collaborative shift scheduling systems},  in
  \emph{Proceedings of the 2020 CHI Conference on Human Factors in Computing
  Systems}, pp.~1--13, 2020,
  \href{https://doi.org/10.1145/3313831.3376656}{DOI}.

\bibitem{abdulkadiroglu_sonmez_1998}
A.~Abdulkadiro\u{g}lu and T.~S\"onmez, \emph{Random serial dictatorship and the
  core from random endowments in house allocation problems},
  \href{https://doi.org/10.2307/2998580}{\emph{Econometrica} {\bfseries 66}
  (1998) 689}.

\bibitem{abdulkadiroglu_sonmez_2003}
A.~Abdulkadiro\u{g}lu and T.~S\"onmez, \emph{School choice: A mechanism design
  approach}, \href{https://doi.org/10.1257/000282803322157061}{\emph{American
  Economic Review} {\bfseries 93} (2003) 729}.

\bibitem{budish_2011}
E.~Budish, \emph{The combinatorial assignment problem: Approximate competitive
  equilibrium from equal incomes},
  \href{https://doi.org/10.1086/664613}{\emph{Journal of Political Economy}
  {\bfseries 119} (2011) 1061}.

\bibitem{budish_cantillon_2012}
E.~Budish and E.~Cantillon, \emph{The multi-unit assignment problem: Theory and
  evidence from course allocation at {Harvard}},
  \href{https://doi.org/10.1257/aer.102.5.2237}{\emph{American Economic Review}
  {\bfseries 102} (2012) 2237}.

\bibitem{kurokawa_leximin_2018}
D.~Kurokawa, A.D.~Procaccia and N.~Shah, \emph{Leximin allocations in the real
  world}, \href{https://doi.org/10.1145/3274641}{\emph{ACM Transactions on
  Economics and Computation} {\bfseries 6} (2018) 11:1}.

\end{thebibliography}\endgroup

\end{document}